\documentclass[fontsize=11pt, paper=letter, twocolumn]{scrartcl}
\usepackage[utf8]{inputenc}
\usepackage[english]{babel}
\usepackage{authblk}
\usepackage{siunitx}
\usepackage{amsmath}
\usepackage{amssymb}
\usepackage{graphicx}
\usepackage[includeheadfoot, headsep=.1in, margin={.6in,.6in}]{geometry}
\usepackage{hyperref}

\begin{document}

\title{3D correlative single-cell imaging utilizing fluorescence and refractive index tomography}

\author{Mirjam Sch\"urmann, Gheorghe Cojoc, Salvatore Girardo, Elke Ulbricht, Jochen Guck\footnote{To whom correspondence should be addressed: jochen.guck@tu-dresden.de}, Paul M\"{u}ller}

\affil{Biotechnology Center of the TU Dresden, Germany}

\maketitle

This is the pre-peer reviewed version of the following article: \textit{3D correlative single-cell imaging utilizing fluorescence and refractive index tomography}, which has been published in final form at \url{https://dx.doi.org/10.1002/jbio.201700145}.

\abstract{
Cells alter the path of light, a fact that leads to well-known aberrations in  single cell or tissue imaging.
Optical diffraction tomography (ODT) measures the biophysical property that causes these aberrations, the refractive index (RI).
ODT is complementary to fluorescence imaging and does not require any markers.
The present study introduces RI and fluorescence tomography with optofluidic rotation (RAFTOR) of suspended cells, quantifying the intracellular RI distribution and colocalizing it with fluorescence in 3D.
The technique is validated with cell phantoms and used to confirm a lower nuclear RI for HL60 cells.
Furthermore, the nuclear inversion of adult mouse photoreceptor cells is observed in the RI distribution.
The applications shown confirm predictions of previous studies and illustrate the potential of RAFTOR to improve our understanding of cells and tissues.
}

\section{Introduction}
The interaction of light with cells plays a central role in microscopic imaging, not only because of imaging aberrations due to diffraction at cells and tissues, but mainly due to the growing amount of novel techniques that depend on the detailed understanding of light-cell interaction \cite{Guck_2001, Kam_2001, Wax_2002, Scarcelli_2015}.
The property that governs this interaction is the refractive index (RI) distribution of a cell.
It is known that the RI is proportional to mass density for biological materials \cite{Barer1952, davieswilkins1952} and thus imaging its distribution directly gives access to structural information \cite{Choi2007, Lue:08}.
The average RI has been used to identify the differentiation state of a cell \cite{Chalut2012}.
By colocalizing the RI with fluorescence images, dense cellular components, such as mitochondria  and lipid droplets, could be identified  as regions with high RI values \cite{Bon2012, Kim2016}.
Knowledge of the RI also allows to characterize the mechanics of cells with the help of photonic techniques.
For instance, the RI is necessary to calculate the optical forces acting in the optical stretcher \cite{Guck_2001} or to convert a measured Brillouin shift to elasticity \cite{Scarcelli_2015, Meng_2016}.
The cell with its RI distribution can be seen as an optical element with a specific function.
For instance, the optical properties of cells and nuclei improve the passage of light through the inverted vertebrate retina  \cite{Solovei2009, Kreysing:10}.
A recent debate in the biophotonics community centers on the question, whether the nucleus of a cell has a lower RI than the cytoplasm \cite{Brunsting_1974, Choi2007, Lue:08}.
Adding to this debate, we were able to show that the RI of isolated nuclei is lower compared to the cell average RI for several cell lines \cite{Schuermann2016}, which was confirmed by an independent study \cite{Steelman2017}.
However, these studies measured the RI of isolated nuclei, raising the question whether the nuclear RI \textit{in situ} is also lower.
These examples illustrate the need for techniques that combine RI and fluorescence imaging in 3D to quantify the RI in specific  intracellular regions.

In general, the RI is determined by measuring its phase-shifting effect on light.
Quantitative phase imaging is commonly achieved using interference-based techniques, such as digital holographic microscopy (DHM) \cite{Schuermann2015} or quadriwave lateral shearing interferometry \cite{Bon_2009}.
A 2D phase image of a cell represents its optical thickness, i.e., RI multiplied by distance.
With a tomographic approach, this phase information can be transformed into a 3D RI distribution with subcellular resolution.

One such tomographic approach is optical diffraction tomography (ODT) which employs the Rytov approximation to correct for diffraction of light at cells \cite{Devaney_1982, Mueller15arxiv}.
ODT requires the acquisition of a complex field sinogram -- a set of phase and intensity images from different directions.
There are several approaches to obtain such sinograms. 
The most common techniques illuminate the specimen from multiple directions through a high NA objective, reaching an angular coverage of up to 140 degrees \cite{Su2013, Choi2007, Debailleul08, KimY:14, Sung2009, Kim2016, Sung2014}.
To record a sinogram with an angular coverage of 180 degrees and above, multiple techniques, including optical tweezers \cite{Habaza:15, Diekmann_2016} and the mechanical rotation of the specimen or the imaging device \cite{Charriere2006, Kostencka2015, Lin2014}, were proposed.
In the present study, we combine a 360 degree optofluidic rotation \cite{Kolb_2014} with phase and fluorescence imaging.
The optofluidic rotation enables us to record complex field and fluorescence sinograms of single cells in suspension in a contact-free manner.

Due to the fact that the sample rotation is continuous, tomographic imaging with optofluidic rotation requires a preprocessing step to register the discrete angular positions of each sinogram image.
In a previous study, we showed that this registration is possible by tracking a distinctive diffraction spot created by a high RI gradient within the cell \cite{Mueller15tilted}.
However, this approach cannot be applied in general, because not all cell types exhibit such prominent features.
In addition, the complex field and fluorescence images need to be aligned in space and time to allow a correct colocalization of the resulting 3D data.
A universal tomographic approach that addresses colocalization and sample rotation with an unknown angular velocity has not been described so far.

Here, we present RI and fluorescence tomography with optofluidic rotation (RAFTOR), including a novel data processing pipeline that allows to correct for spatial drift and average angular velocity changes during sample rotation.
We assess the ODT reconstruction quality of our approach using a phantom specifically designed to mimic RI values of cells.
By colocalizing the 3D RI with the 3D fluorescence data, we are able to resolve and characterize intracellular structures which we showcase by segmenting and quantifying the RI of the nucleus.
For an exemplary HL60 cell, we measure a nuclear RI that is lower than the RI of the cytoplasm, which confirms previous results \cite{Schuermann2016, Steelman2017}.
In addition, a 3D RI analysis of an adult mouse photoreceptor cell shows an increased RI at the center of the cell as a result of terminal differentiation associated with an inversion of the chromatin structure \cite{Solovei2009}.
Thus, RAFTOR provides an accurate  quantification of the intracellular RI distribution in 3D and demonstrates the advantage of complementing ODT with fluorescence tomography for the biophysical characterization of single, suspended cells.

\section{Experimental realization}
\begin{figure}
\includegraphics[width=\linewidth]{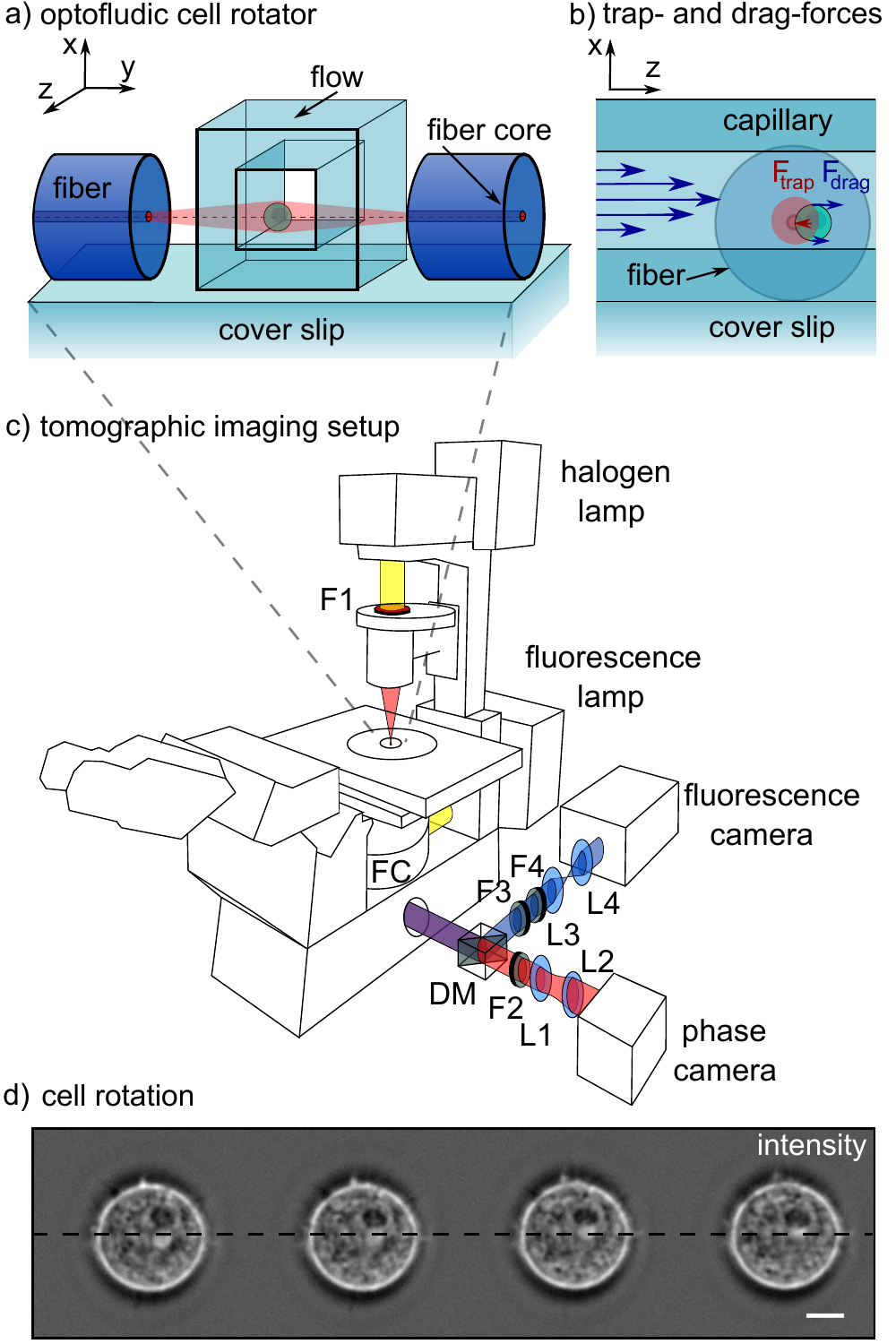}
\caption{Optofluidic cell rotator (OFCR) and imaging setup}{(\textbf{a}) Schematic drawing of the alignment of the OFCR.
Two optical fibers are placed opposing each other with a distance of $\SI{160}{\micro\meter}$ and enable trapping of the cell in 3D. 
A square glass capillary (inner diameter $\SI{80}{\micro\meter}$) is positioned perpendicular to the optical fibers on an alignment SU-8 photoresist mask (not shown).
The center of the optical trap is located $\SI{17.5}{\micro\meter}$ below the center of the glass capillary. 
To rotate the trapped cell (light blue object), a flow is introduced in the channel as indicated by the arrow. 
The side view (\textbf{b}) shows the optical- and drag-forces that act on the cell in the channel (red and blue arrows). 
(\textbf{c}) Schematic drawing of the experimental setup for phase and fluorescence imaging (see text). 
(\textbf{d}) Intensity images of a rotation of an HL60 cell in the OFCR by a total of $\SI{24}{}$ degrees. Scale bar, $\SI{5}{\micro\meter}$. }
\label{fig:OFCR_setup}
\end{figure}

To perform 3D RI tomography, two steps are required: sinogram acquisition and subsequent reconstruction. 
We acquire the sinogram images during contact-free rotation of single, suspended cells using an optofluidic cell rotator (OFCR) (fig. \ref{fig:OFCR_setup}a).
In the OFCR, cells are trapped in three dimensions with a dual-beam laser trap \cite{Lincoln_2007, Lincoln_2007b} built from a fiber laser ($\lambda$\,=\,\SI{780}{\nano\meter}, EYLSA 780, Quantel laser) with two optical fibers (PM780-HP, Nufern).
To induce a rotation of the trapped cell, we introduce a microfluidic flow.
The asymmetric drag forces acting on the cell, induced by placing the trapping position slightly off-center with respect to the parabolic flow profile, lead to a rotation of the sample around the trapping axis (fig. \ref{fig:OFCR_setup}b).
Optofluidic rotation of cells in a similar configuration was introduced by Kolb et al. \cite{Kolb_2014}.

To record sinogram images, the OFCR was combined with a setup enabling simultaneous complex-field- and fluorescence-imaging (the complex-field images will be referred to as phase images in the following for simplicity). 
The OFCR was placed on a standard inverted microscope (IX71, Olympus) which was combined with a custom-built extension that splits the light path for simultaneous fluorescence and quantitative-phase imaging (fig. \ref{fig:OFCR_setup}c). 
For the quantitative phase images, a halogen lamp (TH4-200, Olympus) mounted at the top lamp-port of the microscope was used as light source. The spectrum was restricted by a band pass filter to \SI{647}{\nano\meter}\,$\pm$\,\SI{57}{\nano\meter} (F1, F37-647, AHF analysentechnik). 
For the fluorescence imaging, a mercury lamp (U-RFL-T, Olympus) was mounted on the back lamp-port. 
An excitation filter as well as a dichroic mirror were mounted in a filter cube (FC, filter set 01 , BP 365/12, FT 395, Zeiss) to separate the excitation from the emission light of the fluorophore. 
The emission filter of the filter-set (F3) was mounted separately. 
A 40x microscope objective was used for imaging (NA 0.65, Zeiss). 
The fluorescence emission of the fluorescent-dye used to stain the sample (peak wavelength $\lambda$ \,=\,\SI{461}{nm}) and the light that carries the quantitative phase information ($\lambda$\,=\,\SI{647}{\nano\meter}\,$\pm$\,\SI{57}{\nano\meter}) were coupled out at the side-port of the microscope.

To separate the fluorescence signal from the quantitative phase information, a dichroic mirror with a cut-off wavelength of $\SI{550}{\nano\meter}$ (DM, DMLP550R, Thorlabs) was installed after the side-port. 
The light carrying the phase information was transmitted by the dichroic mirror. 
After the dichroic mirror, a short-pass filter (F2, FESH0700, Thorlabs) blocked the remaining scattered light of the trapping lasers. Additionally, a telescope (L1, LD2060-A $f$\,=\,-\SI{15}{\milli\meter} and L2, LBF254-075-A $f$\,=\,\SI{75}{\milli\meter}, Thorlabs) was implemented to further magnify the image by a factor of five before it was recorded by a commercial phase-imaging camera (SID4-Bio, PHASCIS). 
The phase-imaging camera captured interferograms from which quantitative phase and intensity images were calculated with a proprietary software (SID4-Bio, v.2.1.0.5, PHASICS). 
The fluorescence emission was reflected by the dichroic mirror. 
An additional short-pass filter (F4, FES0550, Thorlabs) and the emission filter of the filter cube (F3, LP 397, filter set 01, Zeiss) blocked the residual reflected light of the halogen lamp as well as the excitation light of the fluorescence lamp. 
The fluorescence emission was detected with a CMOS camera (Marlin F-146, Allied Vision Technologies). 
A non-magnifying telescope built from two lenses with the same focal length (L3, L4, LB1471-B $f$\,=\,\SI{50}{\milli\meter}, Thorlabs) was implemented in front of the camera. 
This telescope enabled to shift the focal plane slightly and therefore allowed to match the focal plane of the fluorescence camera to that of the phase  camera. 
The setup allows a simultaneous recording of fluorescence and complex field sinograms of single cells. 
The four images in Figure \ref{fig:OFCR_setup}d showcase the optofluidic rotation of an HL60 cell by about $\SI{24}{}$ degrees. A full $\SI{360}{}$ degrees rotation of an HL60 cell can be found in supplementary video S1. 
The rotation becomes apparent due to the two prominent diffraction spots that change position with respect to each other.
They are positioned on opposing sides of the cell and hence appear as dark and bright spots.

\section{Sinogram analysis}
\begin{figure*}[h]
  \includegraphics[width=\textwidth]{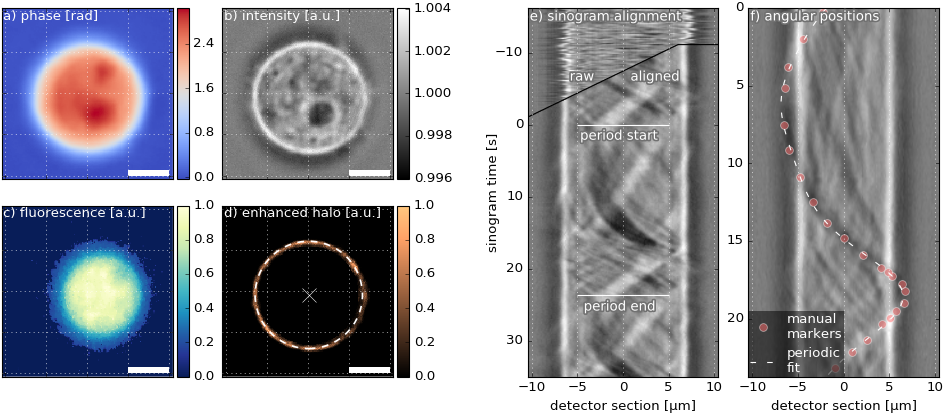}
  \caption{Sinogram alignment and registration of angular positions for a rotating HL60 cell}{
The images show an HL60 cell in (\textbf{a}) phase, (\textbf{b}) intensity, and (\textbf{c}) Hoechst-stained fluorescence. (\textbf{d}) To determine the cell center (white cross) for the sinogram alignment step, the white halo, visible in the intensity image (c), is enhanced (brown ring) and fitted with a circle (dashed line). Scale bar, \SI{5}{\micro \meter}. (\textbf{e}) The two sinograms before (above black line, noisy due to lateral movement) and after (below black line) alignment illustrate necessity and success of the translational alignment step. The registration of the rotational positions (\textbf{f}) is performed for one full rotation (period labeled in (e)) using a sinogram slice that exhibits a sinusoidal signal.
  \label{fig:sino_al}}
\end{figure*}
In practice, optofluidic rotation is prone to translational movements of the imaged object in the optical trap. Additionally, objects often exhibit a rotation with varying angular velocity.
A successful tomographic reconstruction of the 3D RI and fluorescence information requires a correction for these irregular movements in the complex field and fluorescence sinograms.
To address this issue, we employed a novel sinogram analysis pipeline. The phase and intensity images (fig. \ref{fig:sino_al}a,b) were used for the alignment process and the fluorescence images (fig. \ref{fig:sino_al}c) were aligned accordingly by superposition in an additional analysis step.

To correct for the translational movement, we used an object-optimized preprocessing step and a subsequent feature-based algorithm to find the object center as described in detail in section \ref{ap:sino_al}.
For instance, the white halo that is visible in the intensity image of the HL60 cell in figure \ref{fig:sino_al}b can be enhanced (fig. \ref{fig:sino_al}d) to allow fitting of a circle and subsequent alignment of the full sinogram using the circle center positions.
The improvement of the sinogram data quality after such an alignment step is illustrated in figure~\ref{fig:sino_al}e.

To characterize the varying angular velocity during the rotation, we analyzed a sinogram slice that contains visible sinusoidal structures.
These sinusoidal structures are generated by diffraction spots and appear with bright-dark transitions in the intensity sinogram (fig. \ref{fig:sino_al}f). 
For a non-uniform rotation, the sinusoidal structures are skewed but can be traced and fitted with a smooth periodic function (see section \ref{ap:sino_reg}).
The data from the fit allowed us to register an angular position to each of the recorded sinogram images.
These angular positions were then incorporated in the tomographic reconstruction process for both, RI (backpropagation with the Rytov approximation) and fluorescence (backprojection), as described in section \ref{ap:tomo}.
Our sinogram analysis pipeline, implemented in the Python programming language, is available upon request.

\section{Results}

\subsection{Assessment of 3D reconstruction quality}
\begin{figure}
\begin{center}
		\includegraphics[width=\linewidth]{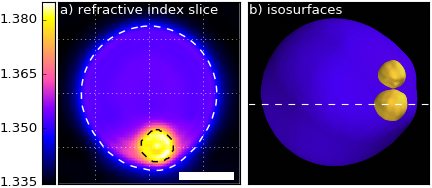}
		\end{center}
	\caption{Reconstruction challenge with cell phantom}{(\textbf{a}) Slice through the 3D refractive index reconstruction of an artificial cell phantom consisting of two silica beads inside a larger hydrogel sphere (see text).
The dashed lines show a segmentation of the phantom and the larger silica bead at $n_\text{white}$\,=\,1.35 and $n_\text{black}$\,=\,1.377 (scale bar, \SI{5}{\micro \meter}). 
(\textbf{b}) 3D isosurfaces of the segmentation. 
The dashed line indicates the position of the cross-section shown in (a).}
	\label{fig:phantom}
\end{figure}
To validate the accuracy of the obtained 3D RI distributions, we analyzed the resolution and the RI with an artificial cell phantom. 
The cell phantom was manufactured using two materials, silica beads (Microparticles, $d$\,=\,\SI{2.79}{\micro\meter}\,$\pm$\,\SI{0.12}{\micro\meter} (mean\,$\pm$\,SD)) embedded in hydrogel spheres (preparation described in \ref{sec:phantom_prep}).

To obtain ground truth values for the RI values of the phantom, we acquired 2D phase images of both building materials individually and determined average RI values by exploiting sphericity with the method described in \cite{Schuermann2015}. 
Hydrogel beads, manufactured in the same way as the phantom, had an average RI of $n_{\text{hydro2D}}$\,=\,\SI{1.356}{}\,$\pm$\,\SI{0.004}{} (mean\,$\pm$\,SD, $N$\,=\,54).
The average RI of the silica beads was $n_{\text{silica2D}}$\,=\,\SI{1.430}{}\,$\pm$\,\SI{0.001}{} ($N$\,=\,28) which we additionally confirmed with RI matching. For the RI measurement of the silica beads we have also considered that silica beads in the small size regime as those used in this study are porous and hence can change the RI when the content of the pores is changing \cite{Garcia-Santamaria_2002, Hu_2014}. 
The data of all 2D phase measurements are shown in supplementary figure S2.

A 3D RI reconstruction of a cell phantom containing two silica beads is shown in figure \ref{fig:phantom}. 
In the isosurface representation (fig. \ref{fig:phantom}b), the two silica beads are clearly visible within the larger hydrogel bead.
The sinogram data and a $z$-stack of the reconstructed 3D RI are shown in the supplementary videos S3 and S4.
The diameters of the two beads are $d_\text{1}$\,=\,\SI{2.9}{\micro m} and $d_\text{2}$\,=\,\SI{2.5}{\micro m}.
We obtained these diameters by computing an effective diameter for two cross-sectional slices and averaging those for each bead.
A comparison of the diameters of the silica beads determined from the 3D reconstruction to the size given by the manufacturer show a good agreement. The first diameter $d_\text{1}$ lies within the given standard deviation and the second diameter $d_\text{2}$ is only slightly smaller.
The average RI of the silica beads and the hydrogel was determined using the segmentation shown in figure \ref{fig:phantom}b to be $n_{\text{silica}}$\,=\,\SI{1.380}{} for the silica beads and $n_{\text{hydro}}$\,=\,\SI{1.354}{} for the hydrogel.
Note that for the silica beads, the RI values from the 2D measurements are higher than the values from the 3D tomographic measurement. In contrast, the values for the hydrogel RI agree well, the RI determined from the 3D measurements lies within the standard deviation of the 2D measurement.
Overall, the higher RI of the silica beads compared to the hydrogel is clearly visible and the structure of the phantom is well-resolved in the 3D reconstruction.

\subsection{Colocalization of RI and fluorescence signal}
\begin{figure*}[t]
\begin{center}
\includegraphics[width=.7\textwidth]{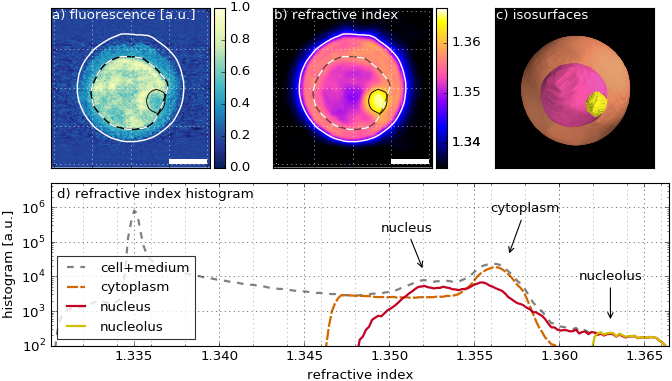}
\end{center}
\caption{Nuclear refractive index is lower than that of cytoplasm}{Slice images (scale bar, \SI{5}{\micro \meter}) of reconstructed (\textbf{a}) fluorescence (Hoechst staining for DNA) and (\textbf{b}) refractive index (RI) illustrate the segmentation of the nucleus (dashed line using fluorescence data) and the cell and the nucleolus (solid lines with $n_\text{white}$\,=\,1.347 and $n_\text{black}$\,=\,1.362). 
The 3D isosurfaces are shown in (\textbf{c}). The log-linear histogram plot (\textbf{d}) shows the different contributions of cytoplasm (cell without nucleus) and the nucleus to the overall RI after segmentation. 
The high RI fraction of the nucleus resembles the nucleolus.  \label{fig:hl60}}
\end{figure*}
To be able to interpret 3D RI maps of cells and to verify the assumption that different organelles have distinct RI signatures, we performed colocalization experiments with tomographic fluorescence imaging. 
Here, we show this colocalization for the nucleus of an HL60 cell.
Fluorescence imaging of the nucleus was realized using the DNA marker Hoechst33342 (Molecular Probes) alongside the RI measurements of the cell. 
The recorded sinograms, which were used for the 3D reconstructions in figure \ref{fig:hl60}, are shown in the supplementary video S1 and had a total of 140 complex field images and 292 fluorescence images.
A slice through the resulting 3D-fluorescence reconstruction of the nucleus and the 3D RI map of the whole cell are shown in figure \ref{fig:hl60}a and \ref{fig:hl60}b for an HL60 cell. 
The complete $z$-stack of the 3D reconstruction of RI and fluorescence are shown in supplementary video~S5.

To determine the different compartments in the 3D reconstruction, we performed segmentation in RI and fluorescence.
A segmentation of the 3D fluorescence map allows a determination of the nuclear shape and position inside the cell, which is indicated by the dashed outline of the nucleus in figure \ref{fig:hl60}a. 
The outline of the complete cell and the high RI region that corresponds to the nucleolus \cite{Fonseca2000} were determined by segmentation of the 3D RI map (fig. \ref{fig:hl60}b).
A 3D representation of the segmentation is visualized by isosurfaces in figure \ref{fig:hl60}c.

Based on the segmentation of the nucleus, the slice through the 3D RI reconstruction in figure \ref{fig:hl60}b shows, apart from the nucleolus, lower RI values in the nuclear area when compared to the cytoplasm. 
The segmentation of the 3D RI map and the correlation with the fluorescence data permits an additional quantification of the RI values in defined regions.
In figure \ref{fig:hl60}d, a histogram of all RI values found in the reconstructed RI map is compared to the histograms of the cytoplasm, nucleus, and nucleolus. 
Note that the main contribution of the cytoplasm has a higher RI value compared to the main contribution of the nucleus. 
The nucleus has a region that has high RI values which we can attribute to the nucleolus. 
Overall, the presented example of RI and fluorescence colocalization tomography show the possibility to determine the RI and thus mass density of specific cellular compartments in their natural environment. 

\subsection{Retina nuclei 3D refractive index distribution}
\begin{figure}[h]
\begin{center}
		\includegraphics[width=\linewidth]{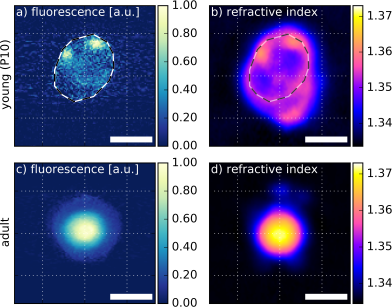}
		\end{center}
	\caption{Refractive index inversion of mouse retina cell nuclei}{Slice images of fluorescence (Hoechst staining for DNA) and refractive index  reconstruction (\textbf{a,b}) of a young mouse retina cell (stage P10) and (\textbf{c,d}) of an adult mouse retina cell (scale bar, \SI{5}{\micro m}). The data showcases the inversion of the nuclear RI after maturation. The position of the nucleus was outlined manually in the fluorescence image (dashed line).}
	\label{fig:retina}
\end{figure}
The RI distribution of cells plays a particularly important role in the inverted retina of vertebrates, where the RI of the tissue influences the light transmission and detection.
Solovei et al. found a curious inversion in the chromatin arrangement in the nuclei of mouse rod photoreceptor cells during development. The finding of this inversion was also the basis for further studies of the light guiding properties of the retina \cite{Solovei2009, Blaszczak2014, Kreysing:10}. 

Using 3D fluorescence and RI colocalization tomography, we observed this inversion also in the 3D RI distribution of mouse retina cells. 
In figure \ref{fig:retina}a and \ref{fig:retina}b a slice through the 3D reconstruction of a retina cell in an early stage of development (P10; 10 days after birth) are shown in fluorescence (Hoechst staining for DNA) and RI reconstruction. 
The corresponding sinogram and $z$-stacks of the reconstruction are shown in the supplementary videos S6 and S7. 
The fluorescence data (fig. \ref{fig:retina}a) is used to outline the nucleus (dashed line) and the outline is transferred to the slice of the RI reconstruction (fig. \ref{fig:retina}b). The RI reconstruction indicates a lower RI in the center of the nuclear region surrounded by a higher RI at the border (fig. \ref{fig:retina}b). 
In contrast, the reconstructed retina cell of the adult mouse (fig. \ref{fig:retina}c,d) is much smaller in size and also shows a more spherical shape (sinogram in video~S8 and reconstruction in video~S9).
In the depicted slice through the 3D RI, the RI is constantly increasing from the rim of the cell towards the center (fig. \ref{fig:retina}d). The same trend is visible for the fluorescence intensity signal. 
This agrees with the mentioned inversion of the chromatin arrangement in mouse rod photoreceptor cells found by Solovei et al. and also with the change in size and shape of the cells described in the same publication \cite{Solovei2009}.
Hence, adult mouse rod photoreceptor cells show an increased RI in the center of 3D RI distribution and therefore act as microlenses \cite{Solovei2009, Kreysing:10, Blaszczak2014}.

\section{Discussion}
The ability to assign RI values to specific intracellular compartments is important for the biophysical characterization of cells. 
We have presented a setup enabling simultaneous phase and fluorescence tomography of single cells that are rotated in a contact-free manner using an optofluidic approach. 
Furthermore, we have presented a novel data processing pipeline which is overcoming deficiencies in the data acquisition such as translational movements of the sample during the rotation or irregularities in the angular velocity.

To assess the quality of the achieved 3D RI maps, we introduced an artificial cell phantom composed of silica beads incorporated in a hydrogel bead. 
The size of the silica beads in the reconstructed 3D RI map showed good agreement with the value of the manufacturer. 
The average RI value found for the hydrogel lies well within the standard deviation determined by 2D phase measurements ($SD$\,=\,0.004) with a difference of only $\Delta n$\,=\,0.002.
The RI values of the silica beads, on the other hand, showed a discrepancy of $\Delta n$\,=\,0.05 which is not within the standard deviation of the 2D measurements ($SD$\,=\,0.001).
This discrepancy has its origin in the data acquisition and reconstruction processes.
The effect of the irregular rotation of the phantom is suppressed by the analysis pipeline but still leads to blurring in the 3D reconstruction (see videos S3 and S4). 
This phenomenon is visible in figure \ref{fig:phantom}a, where the high RI values of the silica bead are not completely confined to a symmetric, circular outline, but appear smeared out.
Due to the resulting mixing of RI values with those of the surrounding hydrogel, the overall RI of the silica bead appears lower compared to the previously determined ground truth values. 
In addition to inaccuracies of the image alignment, the reconstruction can influence the result as well. 
The validity of the Rytov approximation depends on the difference in RI values \cite{Slaney_1984}. 
An RI reconstruction from finite-difference time-domain simulations using the values of the cell phantom (see e.g. \cite{Mueller15odtbrain}) showed that the Rytov approximation underestimates the RI of the silica beads by approximately $\Delta n$\,=\,0.005 (data not shown).
These two phenomena together can explain the lower RI of the silica beads found in the 3D reconstruction compared to the ground truth values.
Nevertheless, the overall trend of a larger RI of the silica beads compared to the hydrogel is still represented correctly in the 3D reconstruction.
The Rytov approximation is valid for sizes and RI values that are common for single cells \cite{Mueller15odtbrain}.
The study of the cell phantom showed that RI values in a regime relevant for the majority of biological samples, i.e. the RI of the hydrogel, is determined correctly by the presented data acquisition and analysis pipeline.

To quantify the nuclear RI of cells, we performed RI and fluorescence colocalization in 3D using RAFTOR.
We could show that the RI of an HL60 cell nucleus is lower compared to the cytoplasm. Thus, previous studies performed with isolated nuclei \cite{Schuermann2016, Steelman2017} are now confirmed \textit{in situ}.
In a second example, we studied the 3D RI distribution within mouse retina cells.
Solovei et al. showed that the chromatin distribution of adult mouse rod photoreceptor cell nuclei is inverted with the more dense heterochromatin localizing at the center and the less dense euchromatin positioned at the border of the nucleus \cite{Solovei2009}.
In the same study, 2D phase images were obtained, which suggested that a higher RI is associated with the heterochromatin regions.
This led to the conclusion that the nuclei of nocturnal mammals with an inverted chromatin architecture should act as microlenses. 
This prediction was theoretically expanded \cite{Kreysing:10} and indirectly confirmed by measuring the focal lengths of different photoreceptor nuclei \cite{Blaszczak2014}. 
Here, we now explicitly measured the increased RI at the center of the inverted adult mouse photoreceptor nucleus for the first time.
We show that the early stage mouse retina cell nucleus has a lower RI at its center with an otherwise irregular distribution, whereas the adult mouse rod photoreceptor cell clearly has a higher RI at the center of the nucleus with a smooth  overall distribution.
Our measurements of the 3D RI agree well with the distributions that were previously used to model light propagation through the inverted retina \cite{Solovei2009, Blaszczak2014}.
Photoreceptor cells of the adult mouse have a small diameter of about $\SI{5}{\micro\meter}$ and high RI values.
As discussed above for the silica beads, the RI of such objects is underestimated by our approach.
Hence, the RI at the center of the mouse photoreceptor cell shown in figure \ref{fig:retina}d with a maximum above $n$\,=\,1.37 could potentially be even higher.
These examples show that colocalization of RI and fluorescence allows to paint an improved picture of the cell.
The knowledge of the RI puts the fluorescence signal into the proper biophysical context.
Conversely, the segmentation of RI by means of fluorescence information allows for a profound interpretation of subcellular regions.

The colocalization of RI and fluorescence in 3D opens up a new dimension in biological imaging, allowing not only to assign biophysical properties to intracellular compartments but also to put flourescently labeled regions or processes into context with their biophysical environment.
As the RI is proportional to mass density, it not only describes optical properties of cells, but also yields information about important biophysical processes \cite{Barer1952, davieswilkins1952}.
For instance, mass density could be used in future studies to accurately describe the process of phase separation undergone by FUS protein droplets \cite{Patel_2015} or by P granules in C.elegans \cite{Brangwynne_2009}.
In addition, the measurement of the cytoplasmic density of eucaryotic cells could help explain phase transitions that are connected to dormancy \cite{Munder_2016}.
The combined knowledge of RI, mass density, and the associated fluorescence information offers new insights to understand the biophysical processes that regulate phase separation \textit{in situ}.

The 3D RI distribution is additionally important for technological advances.
Two examples are Brillouin microscopy and the optical stretcher, optical techniques that are used to characterize the mechanical properties of cells.
In Brillouin microscopy, the RI is essential to compute the longitudinal elastic modulus from the measured Brillouin shifts \cite{Scarcelli_2015}.
Hence, for the computation of a 3D elastic modulus map in Brillouin microscopy, a 3D RI map yields more accurate results than an average RI estimate.
In the optical stretcher, the RI is needed to determine the optical forces acting on the trapped cell \cite{Guck_2001}.
Apart from measuring the RI to study biophysical properties of cells, the 3D RI distribution can also be used to enhance imaging in adaptive optics \cite{Kam_2001}.
For instance, computing the distortions that are caused by a known 3D RI distribution could increase the imaging depth in strongly scattering tissue.
Finally, tomographic imaging in a microfluidic setting, as presented here, allows to image multiple cells in sequence and thus, facilitates cell sorting based on 3D RI and fluorescence.

These examples illustrate the wide range of biological questions and technological advances  that benefit from colocalized 3D RI and fluorescence imaging.
With RAFTOR we now have the tool to address the topics mentioned and to extend biophysical characterization for an improved understanding of cells and tissues.

\section{Materials and Methods}

\subsection{Sample preparation}

\subsubsection{Cell phantom preparation}
\label{sec:phantom_prep}
The cell phantom was produced from silica beads embedded in polyacrylamide (hydrogel). 
The silica beads were commercial beads (Microparticles) with a diameter of $d$\,=\,\SI{2.79}{\micro\meter}\,$\pm$\,\SI{0.12}{\micro\meter} (mean\,$\pm$\,SD) suspended in water. 
The beads were washed with three centrifugation steps in H$_2$O. 
After the last centrifugation step, the supernatant water was removed from the beads and they were immersed in the hydrogel. 
A flow-focussing microfluidic-system was used to produce cell phantoms from the hydrogel silica bead mixture (for a similar system see \cite{Christopher_2007}). The resulting phantoms were suspended in phosphate buffered saline (PBS).

\subsubsection{Cell culture and fluorescence labeling}
Human myelocytic leukemia cells (HL60/S4 referred to as HL60 cells in the text for simplicity) were cultured at $\SI{37}{\degree}$C and $\SI{5}{\percent}$ CO$_2$ in RPMI medium (Gibco) supplemented by $\SI{10}{\percent}$ fetal bovine serum (FBS, Gibco) and $\SI{1}{\percent}$ penicillin-streptomycin. 
For fluorescence imaging, the cells were stained with Hoechst 33342. 
First, cells were centrifuged for $\SI{5}{\min}$ at $115\times g$ and then were resuspended in PBS. 
For the staining, $\SI{0.5}{\micro\liter}$ Hoechst dye was added to $\SI {2}{\milli\liter}$ cell suspension and then incubated for $\SI{20}{\min}$ at $\SI{37}{\degree}$C. 
Subsequently, the cells were washed by centrifugation and resuspended in PBS as described above.
For measurements, the cell suspension was diluted by adding PBS to be able to rotate one cell at a time without flushing a second cell into the field of view.

\subsubsection{Mouse retina cell isolation}
All animal experiments were carried out in strict accordance with European Union and German laws (Tierschutzgesetz). 
All experimental procedures were approved by the animal ethics committee of the TU Dresden and the Landesdirektion Sachsen.
Retina cells were isolated from a mouse at postnatal day 10 (P10) and an adult mouse (C57Bl/6J; Janvier-labs, France). 
Retinae were isolated from excised eyes, incubated in PBS containing $\SI{0.2}{}$-$  \SI{0.4}{\milli \gram \per \milli\liter}$ papain (Roche) for $\SI{30}{\min}$ at $\SI{37}{\degree}$C.
Suspensions were rinsed, spun down and washed with PBS. After a short incubation in PBS supplemented with DNase I ($\SI{200} {U\per\milli\liter}$; Sigma-Aldrich), tissue pieces were triturated with a $\SI{1}{\milli\liter}$ pipette tip to obtain single-cell suspensions.
Cell nuclei were then labeled with Hoechst 33342 as described above.

\subsection{Data acquisition}
Cells were introduced into the microfluidic system with an input and an output reservoir, in which the flow was controlled by adjusting the height of one of the reservoirs. 
A cell was trapped in the OFCR and the focus was adjusted by maximizing the contrast in the phase image and minimizing the contrast in the intensity image.
Subsequently, a flow was induced in the microfluidic channel until a slow but steady rotation of the cell was visible. 
The fluorescence lamp was turned on immediately before the acquisition was started to avoid unnecessary bleaching of the sample. 
The rotation of the cell was recorded on both cameras simultaneously, capturing interferograms with an acquisition rate of $\SI{5.87}{fps}$ and fluorescence images  with $\SI{12.2}{fps}$. 
Recording a full cellular rotation took about $\SI{20}{s}$ to $\SI{60}{s}$.
For each measurement, reference and background images were recorded without the cell but with the laser turned on in exactly the same configuration as during the measurement. 
Furthermore, a corner of the SU8 chip was imaged with both cameras simultaneously after each measurement for a spatial reference between the two imaging systems (see \ref{ap:superpos}).
The RI of the media was determined by an Abbe refractometer (2WAJ, Arcarda).

\subsection{Image preprocessing}
Phase images were background corrected by subtracting a reference phase image and subsequently corrected for background tilt and offset as described elsewhere \cite{Schuermann2015}.
Intensity images were divided by a background intensity image and were subsequently normalized to the average background signal taken from a five pixel border around all sinogram images.
The amplitude was computed by taking the root of the normalized intensity.
Fluorescence images were preprocessed with a total variation denoising filter to inhibit artifacts in the 3D reconstruction.
The background fluorescence signal was determined using an empty region in the fluorescence sinogram and was subsequently subtracted from all fluorescence images.
Finally, a bleaching correction was performed using an exponential fit to the total fluorescence signal computed for each sinogram slice.

\subsection{Superposition}
\label{ap:superpos}
To be able to superimpose the 3D fluorescence and RI reconstruction, we superimposed fluorescence and complex field sinograms temporally and spatially. 
We achieved temporal superposition by comparing time stamps recorded during image acquisition.
To spatially align the sinogram data, we determined the effective resolution of each camera using a USAF test target and located a common feature point in the previously recorded reference images. 
These data allow to accurately map the phase and intensity image coordinates to the fluorescence image data.

\subsection{Sinogram alignment}
\label{ap:sino_al}
To correct for the lateral movement of an imaged object during optofluidic rotation, we made use of feature based alignment algorithms.
To align the four presented samples, we used four different alignment algorithms that are described below.
The aligned images were computed with bivariate spline interpolation according to the obtained lateral offsets.
To avoid image artifacts for complex field data, the interpolation was performed separately for phase and amplitude data.

\subsubsection{Cell phantom}
The cell phantom had a spherical geometry that could be exploited for sinogram alignment.
We applied a Canny edge detection algorithm to the phase sinogram images to reliably determine the circular contour of the cell phantom.
This contour was used for a circle fit to determine the lateral movement during rotation (see supplementary video~S3).

\subsubsection{HL60 cell}
Due to its inhomogeneity, the above method did not work well for the HL60 cell, but served as a good initial approximation for the cell radius $r_0$ and the center $c_0$.
Here, we exploited the fact that the cell exhibited a white halo in the intensity image (fig. \ref{fig:sino_al}b).
The average background amplitude was subtracted from the amplitude image and the result was multiplied by a circular ramp with a radius of $0.8r_0$ located at $c_0$ to lower the contrast at the center of the image.
Then, all values below zero (the dark fraction of the image in fig. \ref{fig:sino_al}b) were set to zero.
Next, the Hough transform was used to determine the best matching integer valued radius $r_1$ and center $c_1$.
The final step was to set all values outside of the circular disk $0.9r_1<r<1.1r_1$ centered at $c_1$ to zero. An exemplary image is shown in figure \ref{fig:sino_al}b.
To determine the actual alignment center, a weighted circle fit was applied to this image (see supplementary video~S1).

\subsubsection{Mouse retina cell at P10}
The mouse retina cell shown in figure \ref{fig:retina}b,c has an irregular shape and thus, the methods above, exploiting spherical symmetry, did not work.
Instead, the cell was aligned using the contour of the phase image, determined with a Canny edge detection algorithm.
The lateral offset for sinogram alignment was then computed from the center of the bounding box of the contour (see supplementary video~S6).

\subsubsection{Adult mouse retina cell}
None of the above methods worked well for the adult mouse retina cell, because the resolution was too low to allow an accurate alignment using its contour.
We applied an iterative approach that is described by the following equation.
\begin{equation*}
d_\mathrm{opt} = \min_{\mathbf{s}} \sum_\mathrm{\mathbf{r}}\left|p_{i,\mathrm{b}}(\mathbf{r}-\mathbf{s}) - p_{0,\mathrm{b}}(\mathbf{r})\right|
\end{equation*}
For each phase image $p_i(\mathbf{r})$, a binary image $p_{i,\mathrm{b}}(\mathbf{r})$ was computed using Otsu's method. This binary image was translated laterally by the vector $\mathbf{s}$ and then subtracted by a reference image $p_{0,\mathrm{b}}(\mathbf{r})$. The sum of the absolute value of the resulting difference image was iteratively minimized by varying $\mathbf{s}$. The optimal overlap $d_\mathrm{opt}$ could be computed with sub-pixel accuracy using bivariate spline interpolation. We used the stopping criterion $|\Delta \mathbf{s}|<\SI{0.01}{px}$ between two consecutive iterations which led to stable alignment results (see supplementary video~S8).

\subsection{Registration of angular positions}
\label{ap:sino_reg}
To determine the angular positions of the object rotation, we used a semi-automatic approach.
Repeating patterns visible along the temporal axis of the recorded phase or intensity sinograms were used to determine the periodicity $T$ of the rotation.
Certain slices through the sinogram, perpendicular to the axis of rotation, revealed sinusoidal curves.
In the intensity sinogram, these shapes are generated by diffraction spots that transition from light to dark as they follow the rotation of the cell.
These sinusoidal curves were traced by manually marking their position in the sinogram. Subsequently, the positions were fitted with a periodic function
\begin{equation*}
p(t) = p_0 + a \, \sin\!\!\left[ \frac{2\pi(t-t_0)}{T} + \sum_{n=1}^{N}\sin\!\!\left( \frac{2\pi b_n n\,t}{T}-c_n\right) \right]
\end{equation*}
with lateral offset $p_0$, amplitude $a$, sinogram time $t$, temporal offset $t_0$, period of the function $T$ (fixed), and the skewing coefficients~$b_n$ and~$c_n$. The skewing sensitivity $N$ must be adjusted according to the degree of skewing and was set to $N$\,=\,2 for the phantom and the HL60 cell sinograms and to $N$\,=\,4 for the retina cell sinograms.
A visualization of the marked positions and the subsequent fit is shown for the HL60 sinogram in figure \ref{fig:sino_al}f. 
The skewing coefficients alone cause a modulation of the pure sine curve and thus enable to characterize object rotations with varying, but periodic, angular velocity.
The curve obtained from the periodic fit was used to assign angular positions to each sinogram slice.

\subsection{Tomographic reconstruction}
\label{ap:tomo}
Due to the fact that the angular velocity of the object in the optofluidic cell rotator cannot be fully controlled during the experiment, the registered angular positions of the complex field and fluorescence sinogram images are not distributed evenly along one full rotation. 
To address this issue, the applied tomographic reconstruction algorithms perform angular weighting of the sinogram images according to the registered angular positions.
RI maps were reconstructed from the aligned sinogram data with a backpropagation algorithm in combination with the Rytov approximation \cite{Kak2001, Mueller15arxiv}. 
Here, we used the Python software library ODTbrain version 0.1.6 \cite{Mueller15odtbrain}. Fluorescence maps were reconstructed with a filtered backprojection algorithm as implemented in the Python library radontea version 0.1.9 \cite{radontea}.

\section{Supplementary materials}
\subsection*{S01\_HL60\_sinogram.avi}
The video shows the aligned phase, intensity, and fluorescence
sinogram images of the HL60/S4 cell.
The displayed rotational angle was determined using a semi-automatic
approach as described in the text. The intensity and fluorescence data is shown in arbitrary units and the phase data is shown in radians.

\subsection*{S02\_2D\_phase\_measurements.pdf}
\textbf{a)} Refractive index (RI) of hydrogel beads determined from 2D phase images $n_\text{hydrogel}$\,=\,1.356\,$\pm$\,0.004 ($N$\,=\,54) measured in phosphate buffered saline (PBS) $n_\text{PBS}$\,=\,1.335.
\textbf{b)}~RI of silica beads measured in a sucrose-water mixture with $n_\text{suc-wat}$\,=\,1.42 resulted in $n_\text{silica}$\,=\,1.430\,$\pm$ \,0.001 ($N$\,=\,28).
Due to the porosity of the silica beads, some buffers are better suited than others for RI measurements.
For instance, a mixture of Thiodiethanol (TDE) and water leads to inconsistencies in the measured RI of the silica beads at different time points after immersion, which could be explained by a varying concentration of TDE within the porous silica beads.
Using a sucrose-water mixture, we did not observe such inconsistencies.
\textbf{c)} RI matching of silica beads using a sucrose-water mixture. 
The RI of the beads is determined from the dark-bright transition, which occurs at RI values of the sucrose-water mixture between 1.4275 and 1.431.
For each condition, an exemplary bead image is shown.
The RI of the silica beads determined from the phase measurement shown in (b) lies within this range.
\textbf{d)} 
To make sure that the porosity of the silica beads does not affect their RI during phantom production, we measured silica beads in hydrogel at two time points.
For the measurements, a thin hydrogel film was produced between two coverslips and had a height of about about \SI{165}{\micro m}.
The hydrogel has the same composition as the hydrogel used for the phantom preparation.
The RI is measured at two different time points, directly after immersing the silica beads in the unpolymerized hydrogel and \SI{30}{min} later.
The \SI{30}{min} correspond to the time it takes to prepare one batch of cell phantoms.
The polymerization of the hydrogel is in both conditions only induced once it is added onto the cover slip.
The RI of the beads is given as change in RI compared to the background $\Delta n$. 
The two measurements show no significant difference when tested with the Mann-Whitney-Test with a significance level of $p$\,=\,0.05, which shows that the porosity discussed in (b) does not affect the RI of silica beads embedded in hydrogel.
In all plots, the box shows one SD and the whiskers show 2\,SD.
The mean is shown as a square and the median as a line.

\subsection*{S03\_phantom\_sinogram.avi}
The video shows the aligned phase and intensity sinogram images of the cell phantom. The displayed rotational angle was computed by assuming
a constant angular velocity.
The intensity data is shown in arbitrary units and the phase data is shown in radians.

\subsection*{S04\_phantom\_3D\_slice\_RI.avi}
The video shows z-slices through the 3D refractive index reconstruction of the cell phantom.
The large and the small glass beads are visible at $z$\,=\,-\SI{1}{\micro m} and $z$\,=\,+\SI{1}{\micro m}.

\subsection*{S05\_HL60\_3D\_slice\_RI\_fluor\_seg.avi}
The video shows the 3D refractive index and fluorescence reconstruction of the HL60/S4 cell using the sinogram shown in supplement 1.
In addition, the segmentation of the nucleus that  was computed from the fluorescence images is shown as binary images.

\subsection*{S06\_mouse\_retina\_p10\_sinogram.avi}
The video shows the aligned phase, intensity, and fluorescence sinogram images of a mouse retina cell at stage P10. 
The displayed rotational angle was determined using a semi-automatic approach as described in the text.
The intensity and fluorescence data is shown in arbitrary units and the phase data is shown in radians.

\subsection*{S07\_mouse\_retina\_p10\_3D\_slice\_RI\_fluor.avi}
The video shows the 3D refractive index and fluorescence reconstruction of the P10 mouse retina cell.

\subsection*{S08\_mouse\_retina\_adult\_sinogram.avi}
The video shows the aligned phase, intensity, and fluorescence sinogram images of an adult mouse retina photoreceptor cell.
The displayed rotational angle was determined using a semi-automatic approach as described in the text.
The intensity and fluorescence data is shown in arbitrary units and the phase data is shown in radians.

\subsection*{S09\_mouse\_retina\_adult\_3D\_slice\_RI\_fluor.avi}
The video shows the 3D refractive index and fluorescence reconstruction of the adult mouse retina photoreceptor cell.

\section*{Acknowledgements}
We thank Katrin Wagner for initial discussion and tests for the cell phantom production. We thank the BIOTEC/CRTD Microstructure Facility (partly funded by the State of Saxony and the European Fund for Regional Development - EFRE) for the production of the hydrogel beads and phantoms.
The HL60 cells were a generous gift from Donald and Ada Olins (University of New England). The authors thank Iain Patten, Corinna Blasse, Maj Svea Grieb and Giorgos Tsoumpekos for helpful discussion of the manuscript. 
This project has received funding from 
the European Research Council Starting Grant ``LightTouch'' (grant agreement number 282060 to J.G.) and from the Alexander-von-Humboldt Stiftung (Humboldt-Professorship to J.G.).

\bibliographystyle{ieeetr}
\bibliography{article}

\end{document}